# Dynamics of electric charge transport and determination of the percolation insulator-to-metal transition in polyvinyl-pyrrolidone / nano-graphene platelet composites

Charalampos Lampadaris[1], Ilias Sakellis[1,2] and Anthony N. Papathanassiou[1, *]

[1] *National and Kapodistrian University of Athens, Physics Department, Solid State Physics Section, Panepistimiopolis, GR 15784 Zografos, Athens, Greece*

[2] *NCNR Demokritos, GR 15341 Agia Paraskevi, Greece*

**ABSTRACT**

Polyvinyl-pyrrolidone (PVP) loaded with different fractions of dispersed nano-graphene platelets (NGP) were studied by Broadband Dielectric Spectroscopy in the frequency range from 1 mHz to 1 MHz. Complex permittivity and dynamic ac conductivity as a function of frequency, temperature and composition were explored. The concentration-dependent insulator-to-conductor transition was traced through dependence of the dc conductivity and the onset of the dispersive ac ac conductivity. The temperature evolution of the dielectric spectra, below and above the fractional threshold exhibits different dynamics and signs the critical percolation threshold. Percolation is dictated by quantum penetration of the effective potential barrier set by the polymer matrix operating in parallel with conduction along physical contact of NGP, in accordance with predictions for systems consisting of a semi-conducting matrix and dispersed conducting inclusions.

* Corresponding author; e-mail address: antpapa@phys.uoa.gr

Functional polymers with dispersed nano-structures exhibit properties of significant technological importance, broad applicability and low production cost. Polyvinyl-pyrrolidone (PVP) is a polymer wide known for its pharmaceutical applications. It can be used as binder, coating and disintegrate for tablets stabilizer [1]. PVP is easily dissolved in water; both PVP and water are non-toxic, environmentally friendly and bio-compatible materials. Graphene is an excellent conductor of electricity and can be dispersed into water. Nano-graphene platelets combine the advantageous properties of graphene and the low production cost (e.g., in relation with single layered graphene). Thus, controlling the PVP/NGP composition, one can tune the properties of the composite, respectively, and achieve the desired optimum properties respectively. In this way, novel ambitious devices can be innovated. For example, water solutions PVP/ NGP composite can be used as bio-compatible link between electrodes and skin or tissues. As a result, various coated functional polymers with graphene [2, 3] have been prepared in order to analyse the mechanisms of charge transfer [4], to find the critical nano-particle concentration to achieve percolation of electric charge carriers along the volume of the composite [5, 6, 7] and generally to inspect the electrical properties for those various matrices. Lastly, PVA (polyvinyl-alcohol) is a relevant material with polyvinyl-pyrrolidone that has been studied significantly [8]. In such studies, the dc conductivity vs composition is examined to find the percolation threshold. Standard percolation theories [9] assume that the host matrix is a perfect insulator, which is an abrupt assumption in many real systems. Recent progresses on percolation phenomena in systems with matrixes that can be penetrated by electron quantum mechanical tunneling [10] describe better PVP/NGP composites.

Aqueous polymer solutions were formed by dissolving PVP K30 (ASG SCIENTIFICCAS 9003-39-8) in de-ionized doubly distilled water and aqueous solutions of nano-graphene platelets (Angstrom Materials N008-P-40 nano-graphene platelets of average lateral dimensions ≤10 μm and average through-plane dimensions ranging from 50 nm to 100 nm) respectively, were separately ultra-sonicated in a heat bath at 345 K for a time duration of 2 hours. Subsequently, the solutions were mixed together and cooled to room temperature under ultra-sonication. The aqueous mixture was drop casted on a clean flat Teflon surface and was dried at ambient conditions for, at least, two days. The upper surface of the solid nano-composite films were grated to obtain a flat surface parallel to the one facing the teflon plate. The thickness of the specimens was a few of millimeters. SEM imaging indicated a uniform dispersion of the nano-particles within the polymer matrix (Fig. 1).

Specimens were cut from the sold nano-composite film, which were placed inside a capacitor type sample holder of a liquid closed-circuit helium cryostat (ROK, Leybold-Heraeus) operating from 15 K to room temperature. Temperature was stabilized with an accuracy of 0.01 K by a LTC 60 temperature controller. Complex permittivity measurements were collected from 1 mHz to 1 MHz with a Solartron SI 1260 Gain-Phase Frequency Response Analyzer, and a Broadband Dielectric Converter (BDC, Novocontrol). WinDeta (Novocontrol) software was monitoring the data acquisition [11, 12].

The real and imaginary parts of the complex permittivity ($\varepsilon'(f)$ and $\varepsilon''(f)$, respectively) constitute the complex dielectric permittivity function $\varepsilon^*$: $\varepsilon^*(f) = \varepsilon'(f) + i\,\varepsilon''(f)$ [13], where $i^2 = -1$ and f is the frequency of an externally applied harmonic field. Alternatively, the t complex electric conductivity $\sigma^*(f) = i\omega\varepsilon_0\varepsilon^*(f)$, consist of a real and an imaginary component $\sigma'(f) = 2\pi f\,\varepsilon''(f)$.

The dielectric spectra of nano-composites of various volume fractions of NGPs depicted in Fig. 2, indicates that the increase of the NGP loading results in a systematic increase of the value of both $\varepsilon'(f)$ and $\varepsilon''(f)$; accordingly, an increase of the conductivity and capacitance, respectively. Augmenting the NGP content, the percolation network is enriched in conducting 'links' and the density of interfaces and, subsequently, charge trapping centers increase, too. The real part of the dynamic conductivity $\sigma'(f)$ of lightly doped composites consists, in a double logarithmic representation, of a low frequency plateau, whereas the frequency independent conductivity coincides with the dc conductivity $\sigma_0$, and a high frequency dispersive region (Fig. 3). We observe that conductivity values increase on volume fraction of NGP. Increasing the NGP content, a percolation network is formed: At high concentrations, the insulating barrier separating the platelets becomes thinner and can be easily penetrated by quantum mechanical tuneling. At higher NGP loadings, in addition to the above mentioned tunneling process, a continuous network is formed due to physical contacting among NGPs.

Percolation theory for systems consisting of conducting inclusions dispersed into an insulating matrix predicts a critical volume fraction $\varphi_C$, below which, a critical value of the volume fraction $\varphi_C$ the composite does not conduct electricity, while, above it, the dc conductivity at constant temperature is given by:

$$\sigma_0(\varphi) \propto (\varphi - \varphi_C)^r \tag{1}$$

where $r$ is an exponent roughly equal to 2 for three dimensional systems [9]. Eq. (1) is valid under the constraint that the host matrix is a perfect insulator impermeable by electric charge carriers. In fact, on increasing the volume fraction of nano-particles, the PVP spacing becomes thinner; the transmission coefficient for quantum mechanical tunneling through the polymer separating neighboring conductive inclusions becomes non-zero. Nano-granular metal composites, where hopping or tunneling between isolated metal grains occurs, are characterized by two distinct electronic transport regimes depending on the relative amount of the metallic phase [10]. On increasing the fraction of the metallic phase, the transition from the dielectric phase to a semi-conducting one is traced, as phonon assist tunneling of electrons through the host material separating metallic regions. At higher concentrations, physical contact of metallic regions provides a continuous network for electric charge flow; this mechanism proceeds in parallel with the mentioned hopping conduction.

The dc conductivity $\sigma_0$, estimated from the low frequency constant ac conductivity data, is plotted against volume fraction at room temperature in Fig. 3. Eq. (1) is insufficient to fit the entire set of data point, i.e., including the data points corresponding to high loading (to about 12 % vol NGP). Reasonable fitting parameters were obtained when eq. (1) was fitted to the low concentration data points, providing $\varphi_C \approx 0.22$ % vol. For volume fraction values sufficiently above the critical one, another transition probably occurs when percolative conduction starts to dominate via physically contacting of NGPs. Due to the high scattering of the experimental points of Fig. 3 on the low loading of NGPs no fitting method could give reasonable parameters. Thus, we provide an approximation at $\varphi_C = 0.2 \pm 0.1\%$ and $r = 1.3 \pm 0.1$. The latter value is comparable with the value 0.5 % vol predicted for ideal insulator/conductor composites. The percolation threshold is quite low in comparison with carbon nanotubes [14], indicating that a very small amount of NGPs, which are considerably cheaper than single layer graphene, can optimize the electrical properties of PVP.

Room temperature σ′(f) plots for composites of varying NGP loading are depicted in Fig. 4. The ac conductivity spectra are characterized by a crossover from the dc plateau to a dispersive high frequency region. The dc conductivity, which is determined from the low-frequency plateau, augments systematically upon NGP content. At significant NGP loading, the spectrum resembles that of a conductor, indication that a continuous percolation network of mutually attaching NGPs has been formed. This picture agrees with that the granular metal composites exhibit [10]. The concentration where the composite behaves as a conductor (and the ac conductivity is loss free) is estimated to be roughly $\varphi'_c \approx 0.33$ %. The threshold 0.33 is consistent with the percolation threshold in three-dimensional continuous network [15, 16, 17]. Although this method seems to be less accurate than others, it has the advantage to have a direct visualization of the dielectric (or semiconducting) state to the conducting one.

BDS measurements were carried out at isothermal conditions below room temperature. Distinct conductivity mechanisms have different temperature dependencies in principle. For example, ionic or hopping conductivity decrease in cooling, while, metallic conductivity has a opposite trend as the vibrational motion of atoms and, subsequently, the effective cross-section for electron back-scattering are suppressed upon freezing. In Fig. 5, isotherms ε′′(f) are demonstrated, respectively, for 0.05, 0.22 and 0.55 % vol NGP. At low NGP fraction, ε′′(f) gets suppressed on cooling (i.e., $\left(\frac{\partial \varepsilon''(f,T)}{\partial T}\right) > 0$), while, for 0.22 and 0.55 % vol NGP, ε′′(f) is a decreasing function of temperature ($\left(\frac{\partial \varepsilon''(f,T)}{\partial T}\right) < 0$). The NGP fraction induces new pathways for electric charge transport and affects both the topology and the energy landscape (i.e., the PVP separating the platelets becomes thinner, the effective potential barrier gets modified and tunneling through it is enhanced. The critical volume fraction signaling the change in the electronic properties of the composite is about $\varphi_C = 0.22$ %, a value close to the one that was found above using the effective minimum approximation model [10]. An increase of the dc conductivity upon cooling at temperatures lower than room temperature has been observed in polystyrene/NGP composites [18]. The decrease of temperature suppresses rotational fluctuation in polymeric molecules; therefore the electric field gains the ability for orientation. The induced polarization of the polymer matrix causes the emergence of opposite charges on the polymer – NGPs interface, enforcing the tunneling through the contact. Polarization of the polymer supports tunneling of free charge carriers (while charges bound to the polymer just keep the electret state). At higher NGP loadings, the obtained effects exist, but they are less pronounced due to decrease of the distance between NGPs, a transition from polarization assisted tunneling to physical contact percolation occurs [18].

In Figure 6, surfaces of the function $\sigma'(f, T)$ are plotted for three different NGP volume fractions: below, close to and above the insulator-to-conductor fractional threshold. The surface corresponding to the conducting composite is characterized by high ac conductivity values, weak suppression upon temperature. Surfaces corresponding to insulating (or semi-conducting) composites exhibit an increase of the ac conductivity values upon temperature.

Isotherms of $\varepsilon''(f)$ vs. f (Figures 2 and 5) are typically described as a sum of a dc conductivity component and a Kohlrausch/Williams/Watts (KWW) function [13]:

$$\varepsilon''(f) = \frac{\sigma_0}{\varepsilon_0 2\pi f^n} + \mathcal{F}[\Delta\varepsilon(t) \cdot \Phi(t)] \quad (2)$$

where $\varepsilon_0$ is the permittivity of free space, $\Delta\varepsilon$ is the intensity of the relaxation mechanism, $\Phi(t) = \exp\left[-\left(\frac{t}{\tau_{KWW}}\right)^{\beta_{KWW}}\right]$ is the correlation function and $\mathcal{F}[\Delta\varepsilon(t) \cdot \Phi(t)]$ denotes the Fourier transform of the function enclosed between the brackets. The $\beta_{KWW}$ parameter ($0 < \beta_{KWW} \leq 1$) is a measure of the asymmetric broadening of the relaxation peak and $\tau_{KWW}$ is the related relaxation time. The broadening of the relaxation peaks is likely to occur, since the motion of one tunneling event affects the next neighboring event one via local elastic strain. We note that the Harviliak-Negami function (HN function) [19] poorly fitted the data points compared with the KWW function. The Grafity software was employed to analyze the dielectric spectra. As can be seen in Fig. 7a, apart from the dc-conductivity component, a couple of KWW relaxations are detected. In Fig. 7(b) the high-frequency dielectric components (appearing as 'knees' in the $\varepsilon''(f)$ spectra) systematically shift towards higher frequency, as temperature is reduced, evidencing for coherent electronic charge flow, within conducting NGP or aggregates [21, 22].

Isotherms of the dc-conductivity $\sigma_0$ evaluated from the complex permittivity data for nono-composites of various NGP loading are depicted in Fig. 8. We observe that the insulator to conductor transition results in the reversal of the curvature of the conductivity plots. Sign reversal of the corresponding slopes occurs, as well, as can be seen in the inset Arrhenius diagrams, for temperatures close to room temperature (RT). Fitting a straight line yields apparent activation energy values $E_{DC,Arrh} \equiv -k(\frac{d\ln\sigma_0}{dT^{-1}})_{T=RT}$, where k is the Boltzmann's constant. Apparent activation energies are given in Table I. While the absolute value of $E_{DC,Arrh}$ is correlated with an effective potential barrier separating neighboring localized states, its *sign* evidences the monotony of the $\sigma_0(T)$ function. Crossing along the insulator to conductor transition upon increasing the NGP fraction, a sign reversal of $E_{DC,Arrh}$ occurs, as an effect of the dynamics of electric charge flow upon the modification of the percolation network. The aforementioned changes are related to the formation of a percolation network consisting of physically contacting NGPs. Percolation along this network dominates FIT, yielding augmented conductivity values in expense of interfacial charge trapping and subsequent capacitance effects and KWW relaxations detected below $\varphi_c$. We notice that, the fluctuation induced tunneling model [20] of electrons through a semi-insulating barrier separating neighboring dispersed conducting nano-particles, was insufficient to fit the conductivity-temperature data. Such failure has been reported in some cases, such as carbon-nano-tubes/poly-epoxy composites [23].

In conclusion, the complex permittivity and dynamic ac conductivity of composites consisting of polyvinyl-pyrrolidone (PVP) and dispersed nano-graphene platelets (NGP) was used to investigate electric charge transport dynamics as the volume fraction of NGP is gradually modified. The insulator to conductor transition was explored by exploiting different dielectric functions and their frequency, temperature and composition dependencies. An alternative version of percolation theory, assuming a semi-conducting matrix, is applied to analyze the dc conductivity vs NGP loading. Alternatively, signature of the crossover concentration was detected by examining the temperature evolution of the ac conductivity spectra, based on the fact that different electric charge transport dynamics (dominating below, close to and above the critical volume fraction for insulator-to-conductor transition) exhibit different temperature dependencies. The effective potential barrier controlling percolation and the average distance between neighboring NGPs that form a percolation network are obtained.

**Table I**

| % vol. NGP | $E_{DC,Arrh}$ (meV) |
|---|---|
| 5.71 | +15±2 |
| 3.36 | +70±10 |
| 2.22 | +30±20 |
| 0.33 | +30±10 |
| 0.22 | -60±10 |
| 0.05 | -30±10 |

*Apparent activation energy values obtained from the Arrhenius plots in the vicinity of room temperature, for samples of various compositions. As explained in the text, the sign reversal indicates the change in the monotony of $\log \sigma_0 (T)$, due to the variation of the topology of the percolation network upon loading. The absolute value of the activation energy correlates to an effective potential barrier, while its sign evidences about the monotony of the dc conductivity upon temperature*

# References

1. F. Haaf, A. Sanner and F. Straub, Polymer Journal 17, 143-152 (1985)

2. D. Galpaya, M. Wang, M. Liu, N. Motta, E. Waclawik and C. Yan, Graphene 1: 30–49 (2012)

3. T. Kuilla, S. Bhadrab, D. Yao, N. H. Kim, S. Bose and J. H. Lee, Progress in Polymer Science 35: 1350–1375 (2010)

4. J. Syurik, O. A. Ageev, D. I. Cherednichenko, B. G. Konoplev and A. Alexeev, Carbon 63 317-323 (2013)

5. V. Mittal, Macromolecular Materials and Engineering 299, 906–931 (2014)

6. K. K. Sadasivuni, D. Ponnamma, J. Kim and S. Thomas, Graphene-Based Polymer Nanocomposites in Electronics (Springer, Berlin, 2015).

7 J. R. Potts, D. R. Dreyer, C. W. Bielawski, and R. S. Ruoff, Polymer 52, 5-25 (2011)

8. I. Tantis, G. C. Psarras, D. Tasis, eXPRESS Polymer Letters 6.283–292 (2012)

9. D. Stauffer and A. Aharony, Introduction to percolation theory: Revised Second Edition (CRC Press, Tel-Aviv, 1994)

10. C. Grimaldi, Theory of percolation and tunneling regimes in nanogranular metal films, Phys. Rev. B 89, 214201 (2014)

11. A. N. Papathanassiou, M. Plonska-Brzezinska, O. Mykhailiv, L. Echegoyen and I. Sakellis, Combined high permittivity and high electrical conductivity of carbon nano-onion/polyaniline composites, Synthetic Metals 209, 583-587 (2015)

12. A. N. Papathanassiou, O. Mykhailiv, L. Echegoyen, I. Sakellis and M. Plonska-Brzezinska, J. Phys. D: Appl. Phys. 48, 285305 (2016)

13. F. Kremer and A. Schönhals, Broadband Dielectric Spectroscopy (Springer, Berlin, 1991), p.59.

14. H. M. Kim, M.-S. Choi, J. Joo, S. J. Cho and Ho Sang Yoon, Phys. Rev. B 74, 054202

15. B. Lorenz, I. Orzall and H. O. Heuer, J. Phys. A: Math. Gen. 26 4711 (1993)

16. R. Consiglio, D. R. Baker, G. Paul and H. E. Stanley, Physica A 319 49 (2003)

17. K. Trachenko, M. T. Dove, T. Geisler, I. Todorov and B. J. Smith, Phys.: Condens. Matter 16 S2623– S2627 (2004)

18. J. Syurik, O.A. Ageev, D.I. Cherednichenko, B.G. Konoplev, A. Alexeev, Carbon 63, 317-323 (2013)

19. F. Alvarez, A. Alegría, and J. Colmenero, Phys. Rev. B 47, 125-130 (1993)

20. P. Sheng, Fluctuation-induced tunneling conduction in disordered materials, Phys. Rev. B 21, 2180–2195 (1980).

21. Rakibul Islam, Anthony N. Papathanassiou,Roch Chan Yu King, Jean-Francois Brun, Frederick Roussel, Appl. Phys. Lett. 107, 053102 (2015)

22. Rakibul Islam, Anthony N. Papathanassiou,Roch Chan Yu King, Frederick Roussel, Appl. Phys. Lett. 109, 4966273 (2016)

23. S. Barrau, P. Demont,† A. Peigney, C. Laurent and C. Lacabanne, Macromolecules, vol. 36 (n° 14). pp. 5187-5194 (2003)

# Figure Captions

**FIG. 1.** SEM images of a PVP/0.03 % vol. NGP composite.

**FIG. 2.** (a): $\varepsilon'(f)$ and (b): $\varepsilon'(f)$ spectra recorded at room temperature for various NGP loadings; from top to bottom: 12.00, 5.71, 0.44, 1.10, 0.27, 0.33, 0.22, 0.55, 0.03, 0.11, 0.05% vol, respectively.

**FIG. 3.** Logarithm of dc conductivity plotted against volume fraction. Eq. (1), which provides the insulator-to-conductor transition, fits the low concentration data points. Another transition from phonon assisted quantum mechanical tunneling conduction to percolation due to physical contact of NGP is observed at high NGP loading.

**FIG. 4.** Dynamic ac conductivity spera $\sigma'(f)$ for PVP/NGP measured in the entire frequency regime from $10^{-2}$ Hz to $10^{6}$ Hz in room temperature for different concentrations from top to bottom: 12.00, 5.71, 0.44, 1.10, 0.27, 0.33, 0.22, 0.55, 0.11, 0.05% vol. respectively. A dashed line has been drawn, for improved visualization of the shift of the crossover frequencies with the increase of NGP loading.

**FIG. 5.** Typical isotherms of $\varepsilon''$ vs f for (a) a non-conducting sample (0.05 % vf of NGP) from top to bottom: 290, 270, 210, 190, 230, 140,110 ,80 K respectively(b) a sample in the transition range (0.22 % vf of NGP) from top to bottom: 294, 285, 275, 255, 235, 215, 115, 155 K respectively -as can be seen from the low frequencies- and (c) a conducting sample (0.55 % vf of NGP) from top to bottom: 90, 50, 16.5, 30, 70, 110, 190, 150, 170, 210, 250, 230, 270, 130, 290 K respectively.

**FIG. 6.** 3D diagrams of the real part of $\sigma'(f,T)$ for three of composites 0.1 % 0.4 % and 6 % vol NGP (from left to right, respectively). Notιψε the different way ac conductivity spectra evolves upon temperature (i.e., the change of the sign of the derivative $\left(\frac{\sigma'(f,T)}{\partial T}\right)$, on increasing φ). The above mentioned sign reversal signatures the transition from the insulating state to the conducting one.

**FIG. 7.** (a) Fitting results of a typical dielectric spectra by using one dc conductivity component and two KWW peaks constitute the dielectric spectra at T=50 K of 0.33 % vol NGP composite. (b) Isotherms recorded at temperature ranging from: 16, 50, 90, 130, 170, 210, 255, 265, 275, 285, 295 K from top to bottom curve, respectively. Cooling suppresses the dc component, in accordance with the predictions for granular metal electronic transport via fluctuation induced tunneling.

**FIG. 8.** Dc- conductivity vs temperature for nano-composites loaded with 5.71, 3.36, 2.22, 0.33, 0.22, 0.05 % vol NGP (top to bottom, curves respectively).The insulator to conductor transition is accompanied by reversal of the sign of both the slope and the curvature. Inset: Typical Arrhenius diagrams in the room temperature limit for composites: 5.71, 3.36, 2.22, 0.33, 0.22, 0.05 % vol NGP (from top to bottom, respectively). Straight lines are best fits of the Arrhenius law to the data points at the room temperature limit.

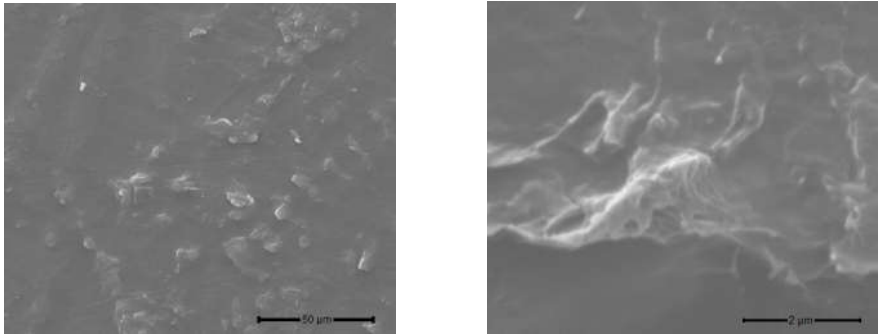

**FIG. 1**

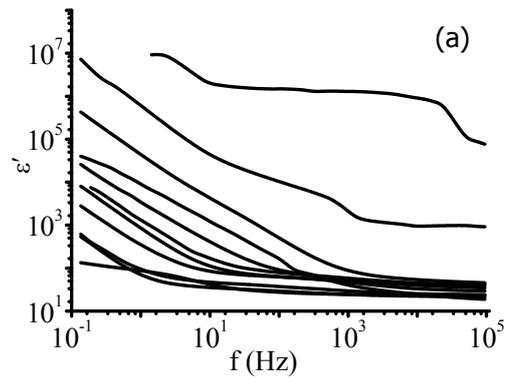

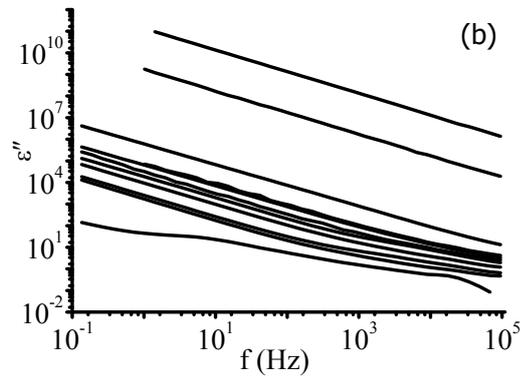

**FIG. 2**

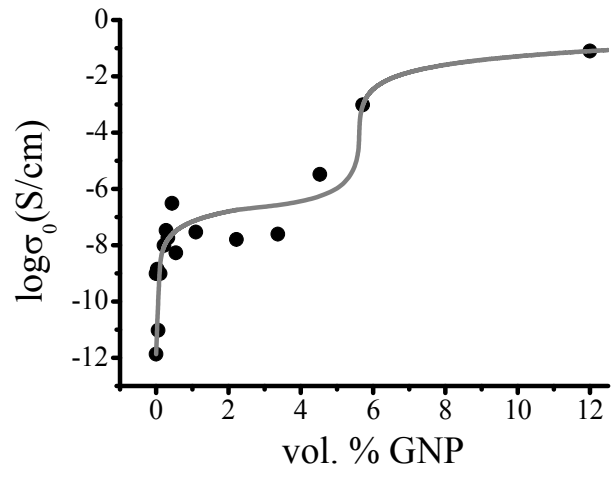

**FIG. 3**

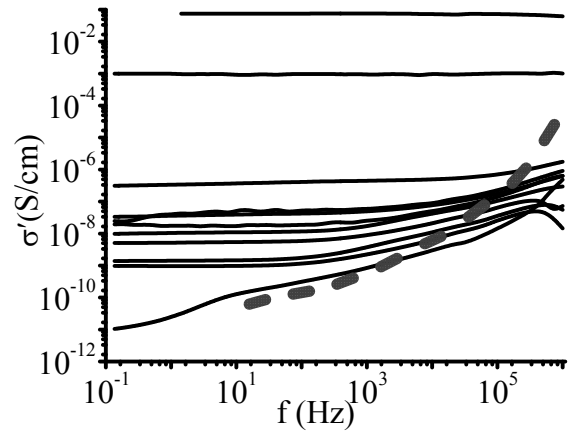

**FIG. 4**

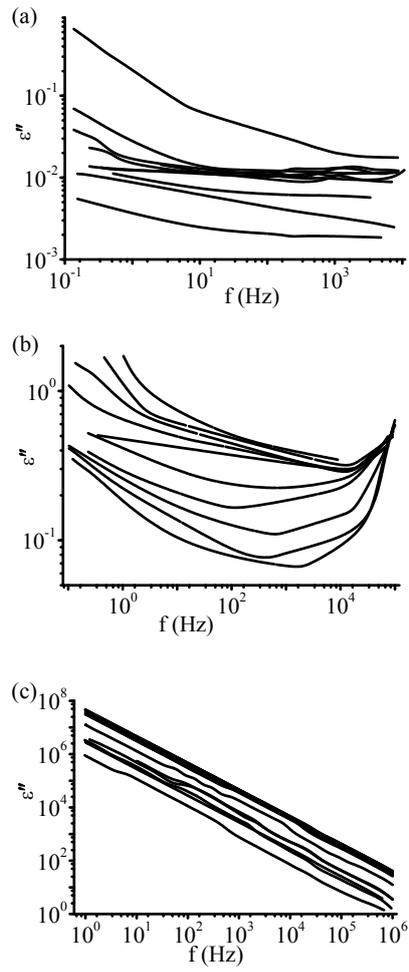

**FIG. 5**

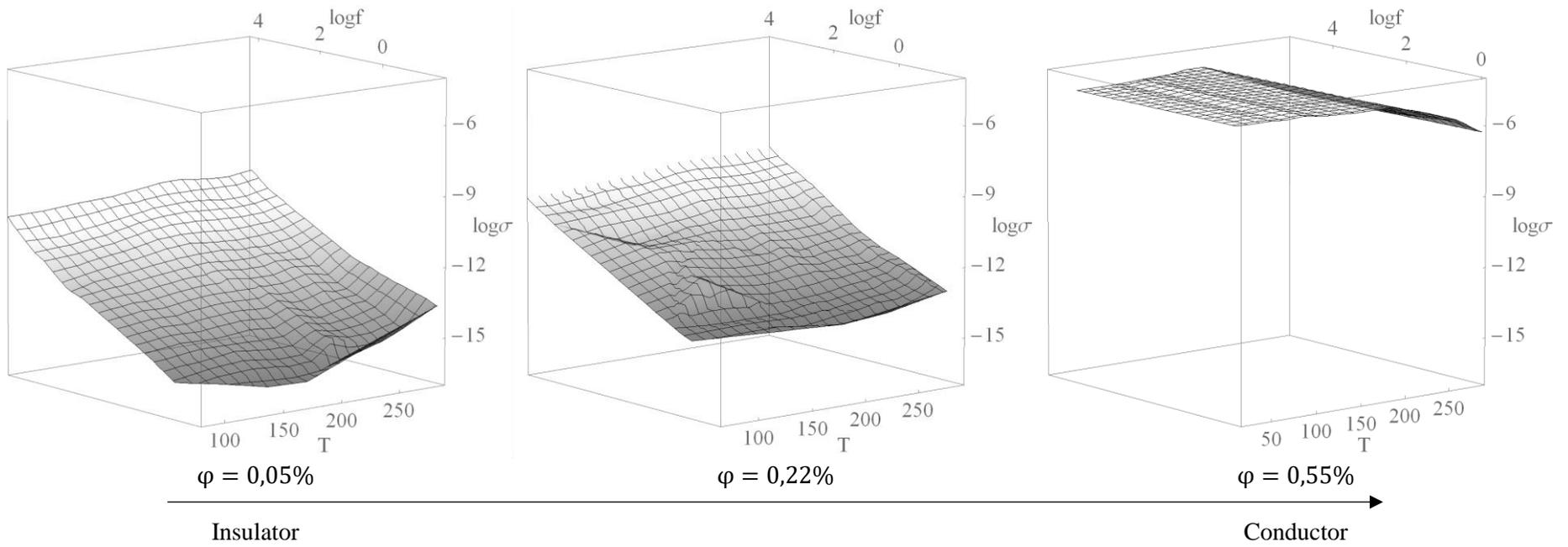

φ = 0,05%  φ = 0,22%  φ = 0,55%

Insulator  Conductor

**Figure 6**

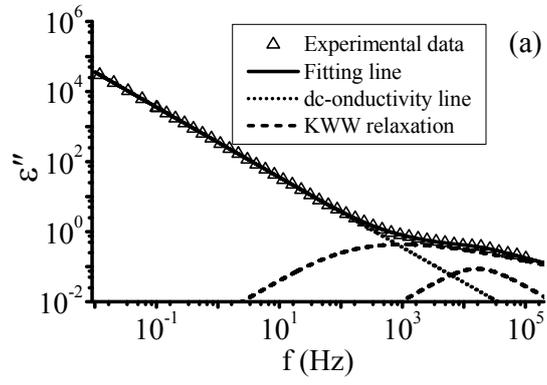
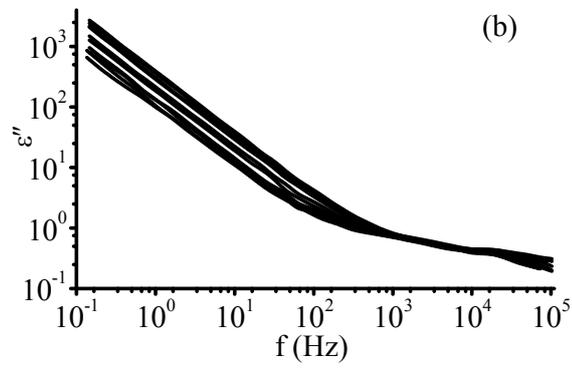

**FIG. 7**

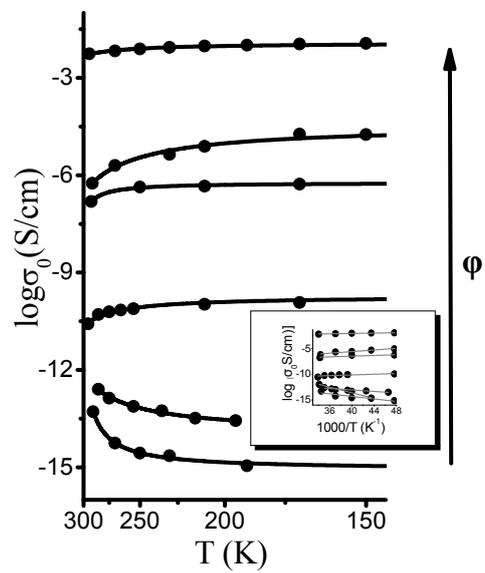

**FIG. 8**